\documentclass[fleqn,twoside]{article}
\usepackage{espcrc2}
\usepackage{graphicx}
\usepackage[figuresright]{rotating}
\newcommand{\lesssim}{\mathbin{\lower 3pt\hbox
   {$\rlap{\raise 5pt\hbox{$\char'074$}}\mathchar"7218$}}} %< or of order
\newcommand{\gtrsim}{\mathbin{\lower 3pt\hbox
   {$\rlap{\raise 5pt\hbox{$\char'076$}}\mathchar"7218$}}} %> or of order

\newcommand{\arcmin}{\ensuremath{{}^\prime}}

\title{The Exploration of the Unknown}
\author{P.~N.~Wilkinson\address[JBO]{University of Manchester, Jodrell
	Bank Observatory, Macclessfield, Cheshire SK11~9DL, United Kingdom; pnw@jb.man.ac.uk}
	K.~I.~Kellermann\address[NRAO]{National Radio Astronomy
	Observatory, 520 Edgemont Road, Charlottesville, VA
	22903-2475, USA; kkellerm@nrao.edu}
		\thanks{The National Radio Astronomy Observatory is
		operated by Associated Universities Inc., under a
		cooperative agreement with the National Science Foundation.}
	R.~D.~Ekers\address[CSIRO]{CSIRO, Australia Telescope National
	Facility, \hbox{P}.\hbox{O}.~Box~76, Epping, NSW 1710, Australia; Ron.Ekers@csiro.au}
	J.~M.~Cordes\address[CU]{Cornell University and NAIC, Ithaca,
	NY, 14850, USA;	cordes@astro.cornell.edu}
		\thanks{This work was supported by NSF grants to
		Cornell University, AST~9819931, AST~0138263, and
		AST~0206036 and also by the National Astronomy \&
		Ionosphere Center, which operates the Arecibo
		Observatory under a cooperative agreement with the \hbox{NSF}.}
	T.~Joseph~W.~Lazio\address[NRL]{Naval Research Laboratory,
		4555 Overlook Ave.~\hbox{SW}, Washington, DC, USA; Joseph.Lazio@nrl.navy.mil}
		\thanks{Basic research in radio astronomy at the NRL
		is supported by the Office of Naval Research.}}
\begin{document}

\begin{abstract}
The Square Kilometre Array is conceived as a telescope which will both
test fundamental physical laws and transform our current picture of
the Universe. However, the scientific challenges outlined in this book
are today's problems---will they still be the outstanding problems
that will confront astronomers in the period 2020 to 2050 and beyond,
when the SKA will be in its most productive years? If history is any
example, the excitement of the SKA will not be in the old questions
which are answered, but the new questions that will be raised by the
new types of observations it will permit.  The SKA is a tool for
as-yet-unborn users and there is an onus on its designers to allow for
the exploration of the unknown. We outline a philosophy for the design
and operation of the SKA that can lead the radio astronomers in the
$21^{\mathrm{st}}$ century to add to the many discoveries of new
phenomena made by radio astronomers in the $20^{\mathrm{th}}$ century.
\end{abstract}

\maketitle

\section{Prologue}\label{sec:eou.intro}

\begin{quote}
\emph{``Now my own suspicion is that the Universe is not only queerer than we suppose,
but queerer than we \emph{CAN} suppose'': J.~B.~S.~Haldane}
\end{quote}

Most of the phenomena we observe today, using telescopes to observe
across the electromagnetic spectrum, were unknown a few decades ago
and, to an amazing extent, were discovered by radio astronomers using
increasingly powerful instruments and either looking for something
else or just following their curiosity. Examples include non-thermal
radiation, radio galaxies, quasars, the cosmic microwave background,
cosmic evolution, pulsars, gravitational lensing, cosmic masers,
molecular clouds, dark matter, and extrasolar planetary systems. These
discoveries have changed astronomy in fundamental ways.  Some
discoveries resulted from increased sensitivity, others from better
spatial or temporal resolution, still others by observing in a new
wavelength band or even testing misguided theory.  Most involved
recognizing a new phenomenon and being able to distinguish it from a
spurious instrumental response.  This scenario is, of course, not
restricted to radio astronomy.  Perhaps the most spectacular example
from astronomy in other wavebands was the discovery of $\gamma$-ray
bursts by a military satellite---currently a major field of
contemporary astrophysics.

It is fashionable to imagine that all research follows some classical
model of the scientific method---formulation of a model or hypothesis
followed by experimental confirmation. Observations not based on
testable theoretical predictions are sometimes called ``butterfly
collecting'' or appeals to ``serendipity'' rather than ``real
science.''  Time allocation committees, referees of grant
applications, and reviewers of instrument proposals tend to focus on
specific questions that will be answered.  Yet, astronomy is not an
experimental science. We can only observe our Universe and its content
with ``eyes'' as wide-open as possible.  We cannot make little changes
or experiments to see what happens, except perhaps for some areas of
planetary research. So how do we plan for discovery? Despite the
apparent capriciousness of our aim, history tells us that a basic
requirement is to carry out systematic work with at least an order of
magnitude improvement over what has been achieved before in one or
more observing capabilities (sensitivity; spatial, temporal, or
spectral coverage; spatial, temporal, or spectral resolution). An
observing instrument which can offer major advances in several
dimensions of parameter space is more likely to make transformational
discoveries---history also shows that much greater sensitivity along
with flexibility of operation is a wise path to follow. The
sensitivity of the Arecibo telescope and the imaging capabilities of
array telescopes are excellent paradigms.

Merely providing access to new areas of parameter space with new
technology is not a sure-fire recipe for making ground-breaking
discoveries, however.  There are other, human, factors to take into
account which are just as important for ensuring the SKA's success as
a discovery instrument. We are all familiar with what is now the
traditional method of using a large common-user telescope involving:
i)~a proposal to tackle a single small problem; ii)~review by time
allocation committees; iii)~the award of a few hours or maybe even
days of observing time; iv)~the analysis of the data via a standard
suite of software, and, then if all goes well, v)~a publication
following filtering by a referee. We dub this ``the standard model''
of observational astronomy and it is perhaps inevitable that the SKA
will allocate much of its operations to this analytic mode.  When,
however, the phenomenon or problem is less well-defined, there can be
a rich mix of possible ``solutions,'' only some of which may have been
explored by theorists and for which the standard model provides a poor
response. It is, therefore, vital to develop a philosophy of design,
operations, and data archival which allows individuals, small groups,
and larger communities freedom to innovate and encourages users to
explore completely new ways of collecting, reducing and analyzing
data---in other words \emph{to allow for discovery as well as explanation}.

\section{The Lessons of Astronomy History}\label{sec:eou.history}

In his 1981 book \textit{Cosmic Discovery} \cite{h81} and in
subsequent articles, Harwit has addressed the question of what factors
lead to new discoveries in astronomy. He argues that a large fraction
of the discoveries have been associated with improved coverage of the
electromagnetic spectrum or better resolution in the angle, time, or
frequency domain. He also notes that astronomical discovery is often
closely linked to innovative new technology introduced into the field
from outside, often from the military. Consequently, many major new
findings have come about more by luck than through careful
planning---although what constitutes ``luck'' is an arguable point
that we discuss in~\S\ref{sec:eou.human}. Nonetheless theoretical
anticipation has usually had little to do with astronomical
discovery---what matters most is the implementation of powerful new
observing tools.

Will progress at the rate achieved in the second half of the
$20^{\mathrm{th}}$ century be likely to continue?  Harwit~\cite{h81}
has tackled this seemingly impossible question in two ways. First by
estimating the fraction of observational phase space which has
presently been explored and then by comparing the number of
discoveries attributable to improved instruments with the number
independently rediscovered, often by totally unanticipated means, with
instruments of quite different kinds. His analysis suggests that we
have already seen perhaps 30\% to~ 40\% of all the major astrophysical
phenomena that will ultimately be revealed by photons, cosmic rays,
neutrinos, and captured extraterrestrial material. While one may be
sceptical about the quantitative accuracy of this prediction,
qualitatively we do not doubt that the Universe still holds plenty of
surprises.

In the first half of the $21^{\mathrm{st}}$ century, powerful tools in
two completely new observational regimes, neutrino and
gravitational-wave astronomy, will become available, and it is very
likely that they will reveal genuinely new phenomena.  Nonetheless,
photon astronomy is far from exhausted, and the low energies of radio
photons and relative ease with which they are generated and propagate
mean that sensitive telescopes in the radio regime will surely
contribute their share of new discovery and understanding.

Moreover, radio observations probe a wide range of conditions---from
dense gasses to dilute, highly relativistic plasmas---are sensitive to
magnetic fields, and yet are not affected by absorption from dust.
The fundamental (baryonic) element of the Universe, hydrogen, has a key
transition at centimetre wavelengths (the 21-cm hyperfine transition).
Radio telescopes routinely make the highest angular resolution
observations in astronomy.  These capabilities have already been
exploited to study some of the most extreme conditions known, e.g.,
the strong gravitational fields in binary pulsars. It is no surprise
that the Key Science Projects currently identified for the SKA exploit
all of the above advantages, and we consider it likely that any future
discoveries---be they from photons, neutrinos, or gravitational
waves---will require radio observations to understand them.

Astronomy at radio wavelengths is marked by a number of differences
from shorter wavelength observations, differences that make radio
astronomy a powerful technique for observing the sky:
\begin{itemize}
\item The sky is mostly empty, which allows unfilled apertures (i.e.,
interferometers) to operate;
\item Long coherent integrations are possible;
\item Large numbers of photons are collected so that the signal can be
amplified and split without any loss in sensitivity; and
\item Diffraction-limited imaging can be obtained via post-processing
so that adaptive optics requires no moving parts.
\end{itemize}

Table~\ref{tab:eou.discovery} lists some of the key discoveries from
radio astronomy in the metre and centimetre wavebands and indicates
the telescopes and the enabling new parameter space. In addition to
adding weight to Harwit's~\cite{h81} emphasis on the importance of
exploiting new technology, several more specific lessons can be
learned.

\onecolumn
\begin{sidewaystable}
\caption{Key Discoveries that Illustrate Discovery Space in Radio Astronomy$^{\sharp}$\label{tab:eou.discovery}}
\begin{center}
\begin{tabular}{|l|l|l|l|}
\hline
&&&\\
\hfil {\bf Discovery} & \hfil {\bf Date} & \hfil {\bf Enabled By$^{\flat}$} & \hfil {\bf Telescope} \\
&&&\\
\hline
Cosmic radio emission 		& 1933	& $\nu$ 
						& Bruce Array (Jansky) \\
\hline
Non-thermal cosmic radiation 	& 1940 	& $\nu$ 
						& Reber antennas \\
\hline
Solar radio bursts		& 1942	& $\nu, \Delta t$ 	
						& Radar antennas \\
\hline
Extragalactic radio sources	& 1949	& $\Delta \theta$
						& Australia cliff interferometer \\  
\hline
21 cm line of hydrogen		& 1951	& theory, $\Delta\nu$ 	
						& Harvard horn antenna \\
\hline
Mercury \& Venus spin rates	& 1962,1965 & radar 	
						& Arecibo		\\ 
\hline
Quasars				& 1962	& $\Delta\theta$
						& Parkes occultation \\
\hline
Cosmic Microwave Background	& 1963  & $\Delta S$, calibration, *theory	
						& Bell Labs horn \\
\hline
Confirmation of General Relativity
				& 1964	& theory, radar, $\Delta t$, 	
						& Arecibo, Goldstone, VLA, VLBI \\
\quad (time delay + light bending) & 1970s & $\Delta\theta$ &\\
\hline
Cosmic masers			& 1965  & $\Delta\nu$ 
						& UC Berkeley, Haystack \\
\hline
Pulsars				& 1967	& $\Omega$, $\Delta t$ 
						& Cambridge 1.8 hectare array \\
\hline
Superluminal motions in AGN	& 1970  & $\Delta\theta$, *theory 
						& Haystack-Goldstone VLBI \\
\hline
Interstellar molecules and GMCs	& 1970s & theory, $\nu, \Delta \nu$ 	
						& NRAO 36-ft \\
\hline
Binary neutron stars
 + gravitational radiation 	& 1974-present 
					& $\Omega$, $\Delta t$, theory 	
					& Arecibo \\
\hline
Gravitational lenses		& 1979	& $\Delta\theta$, theory	
					    & Jodrell Bank interferometer \\
\hline
First extrasolar planet system 	& 1991	& $\Omega$, $\Delta t$ 							& Arecibo \\
\hline
Size of GRB fireball		& 1997	& $\lambda\lambda$, $\Delta S$, theory & VLA \\
\hline
\multicolumn{4}{l}{$\sharp$ This is a short list covering only 
  metre and centimetre wavelengths.} \\
\multicolumn{4}{l}{$\flat$ 
	$\nu$ $\Rightarrow$ spectral coverage;
	$\Delta \nu$ $\Rightarrow$ spectroscopic resolution;
	$\Delta S$ $\Rightarrow$ sensitivity;
	$\Delta t$ $\Rightarrow$ short time resolution.
} \\ 
\multicolumn{4}{l}{~
	$\Omega$ $\Rightarrow$ survey with ample sky coverage.
	$\Delta\theta$ $\Rightarrow$ angular resolution,
	FoV $\Rightarrow$ field of view,
	$\lambda\lambda$ $\Rightarrow$ guided by multiwavelength observations;
} \\ 
\multicolumn{4}{l}{~
	``theory'' $\Rightarrow$ theory played a role in motivating 
         discovery or its search space.
} \\
\multicolumn{4}{l}{~
	``*theory'' $\Rightarrow$ phenomenon was predicted but 
         discovery was independent of the prediction.
} 
\end{tabular}
\end{center}
\end{sidewaystable}

\twocolumn

\begin{itemize}
\item \textbf{Discoveries with radio telescopes have set a large part of the
current astronomical agenda and radio telescopes are now studying
largely what they themselves discovered.}

\item \textbf{The majority of the discoveries (11/17) were \emph{not} a direct result of theory.}
Although there were previous theoretical predictions in two cases,
they played no role in the observational discovery.

\item \textbf{The largest radio telescopes of their day (of a wide range of
types) have dominated the discoveries.}  This
contrasts with Harwit's conclusion that (mainly optical) telescope
size was not a major determinant for success.  There are several
reasons for this difference. Most discrete radio sources are weak,
continuum-only emitters.  Thus, large radio telescopes, which combine
sensitivity and angular resolution, are needed to detect them and to
study their characteristics. This contrasts with the situation at
optical wavelengths, for which even modest-sized telescopes can
observe the myriad of stars with their rich spectroscopic properties.
Moreover, the sensitivity of optical telescopes is often limited by
photon statistics, so it increases only as the square root of the area of the
aperture.  For a radio telescope working in the Raleigh-Jeans part of
the spectrum, sensitivity scales linearly with aperture.  It is
notable that there are no filled aperture radio telescopes less than
64~m diameter in great demand at centimetre wavelengths.

\item \textbf{What a radio telescope was \emph{built for} is almost
never what it is \emph{known for}.}  Almost invariably, the
discoveries in Table~\ref{tab:eou.discovery} were not, often could not
have been, in the minds of the designers of those telescopes.  For
example, Jodrell Bank was built to study meteor trails, Arecibo to
study the ionosphere, and the WSRT to do weak source
counts.  Table~\ref{tab:eou.vla} shows that the \hbox{VLA}, which is
one of the most productive astronomical telescopes of all time, spent
only a quarter of its time during its initial decade of operation on
the key science drivers listed in the funding proposal.

\item \textbf{General-purpose telescopes now dominate the
discoveries.}  While special-purpose instruments dominated discoveries
for the first 30~years, the majority of discoveries since then have
been made with general purpose telescopes---large filled aperture and
arrays of dishes; these are versatile instruments whose performance
can be upgraded by new receiving and signal processing capabilities.
This trend follows the move to ``Big Science'' \cite{p63} as the cost
of facilities with enough sensitivity to continue the exponential
growth required for healthy science becomes too expensive for small
specialized groups.  This lesson encourages us to look for ways in
which operational versatility can be built into an inherently
common-user instrument like the \hbox{SKA}.
\end{itemize}

\begin{table}
\caption{Distribution of VLA Science\label{tab:eou.vla}}
\begin{center}
\begin{tabular}{lc}
\noalign{\hrule\hrule}
Topic & Observing Time \\
\noalign{\hrule}
Stars                   & 16\% \\
Galaxies                & 14\% \\
\textbf{Radio Galaxies} & 13\% \\
\textbf{Quasars}        &  9\% \\
Star formation          &  9\% \\
Solar system            &  6\% \\
AGN                     &  5\% \\
\textbf{Cosmology}      &  4\% \\
Interstellar medium     &  4\% \\
Supernovae              &  4\% \\
Galactic Centre         &  3\% \\
Molecules               &  3\% \\
VLBI                    &  3\% \\
Pulsars                 &  2\% \\
X-ray, etc.             &  1\% \\
Astrometry              &  1\% \\
\noalign{\hrule}
\end{tabular}
\end{center}
\parbox{0.47\textwidth}{Items in bold were key scientific drivers
in~1967 funding proposals.  List compiled by RDE during his tenure at
the \hbox{VLA}.}
\end{table}

\section{New Technology for the SKA}\label{sec:eou.technology}

It is
well established that most scientific advances follow technical
innovation. De~Solla Price~\cite{p63} reached this conclusion from his
application of quantitative measurement to the progress of science
across all disciplines.

Harwit~\cite{h81} pointed out the most important
discoveries in astronomy often result from technical innovation with
the discoveries peaking soon after new technology appears, usually
within~5~years.  However, as a field matures, more general purpose
instruments have more impact.  During the first 30~years of radio
astronomy's brief history, discoveries followed technical innovation,
but we are rapidly approaching the limits of obtaining increased
capabilities simply by upgrading existing telescopes.

De~Solla Price also showed that the normal mode of growth of science is
exponential.  Historical examples included the rate of discovery of
elements and the number of universities founded in Europe.  Some more
recent examples of exponential growth are particle accelerator beam
energy, Internet hosts, and the now famous ``Moore's Law'' for computing
devices.  Such exponential growth cannot continue indefinitely without
a reorganisation or change in technology.

\begin{figure}
\includegraphics[angle=-90,width=\columnwidth]{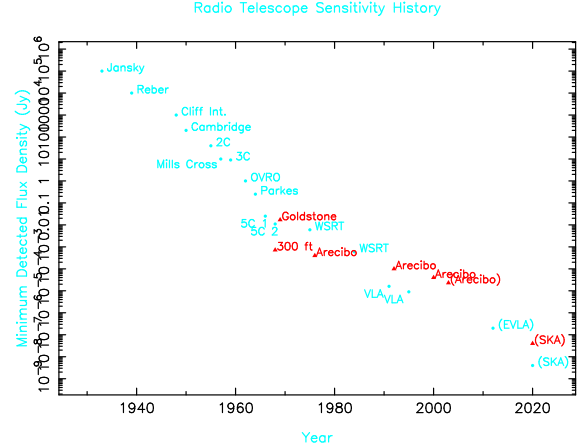}
\vspace*{-1cm}
\caption[]{The history of the weakest radio objects detected using
radio telescopes in both imaging (radio sources, circles) and
time-domain (pulsars, squares) experiments.  The points for the EVLA
and the SKA are projected based on the expected thermal noise limit
for a 100-hour integration.}
\label{fig:eou.time}
\end{figure}

Figure~\ref{fig:eou.time} shows the weakest detected radio sources as
a function of time from the start of radio astronomy before World
War~II to the present day and beyond.  Exponentially increasing
sensitivities have produced an improvement in the weakest sources
detected, for both continuum sources and pulsars, by a factor of
roughly $10^6$ over the past 60~years.  These exponential improvements
in the detection of weak sources have come from a series of technical
innovations involving a combination of increasing collecting area,
decreasing noise temperature, increasing bandwidth, and development of
algorithms for wide-field, high-dynamic range imaging.

Can we continue the exponential growth?  Do we have a new technology?
Receivers are operating near their fundamental sensitivity limits,
bandwidths comparable to the observing frequency will soon be
achieved, time-bandwidth products of order unity are being analyzed,
and interferometry has exploited antenna separations comparable to the
Earth's diameter.  For radio astronomy, the technical challenge is now
to obtain increased collecting area at a reasonable cost, particularly
so for observations of spectral line emission where an increased
bandwidth has no effect on the sensitivity.  Thus, the original
arguments for a square kilometre aperture SKA based on detecting and
imaging the 21-cm line radiation of atomic hydrogen at high redshift
remains.

The combination of transistor amplifiers and their large scale
integration into complex systems which can be duplicated inexpensively
provides one of the keys for change.  The other key technology is the
computing capacity to apply digital processing at high bandwidth,
thereby realizing processes such as multiple adaptive beam formation
and active interference rejection in ways not previously conceivable.
The SKA is very demanding of new technology but the SKA vision has
unleashed an unprecedented burst of creativity in the radio astronomy
and engineering communities.

Technologies that must be developed to realize the SKA include i)~wider bandwidth radio frequency receivers and their direct integration
into antennas; ii)~low-cost manufacturing of large collector systems;
iii)~low-cost signal digitization and conditioning; iv)~radio
frequency interference (RFI) mitigation techniques; v)~low-cost
computing for real-time beam-forming and data processing; vi)~wide-band data transmission via optical fibres; vii)~high speed
real-time imaging of massive data sets, and viii)~applying emerging
computational tools for data mining to the huge data sets that will be
generated.  In contrast to the previous and current generation of
receiver and signal processing equipment, which uses special-purpose
hardware, the next generation will inevitably exploit the convergence
of radio and digital computing technologies---replacing hardware with
firmware or software and allowing unprecedented versatility via the
use of programmable processing engines.

More than simply a quantitative increase in collecting area, though,
the application of these new technologies promises qualitatively new
avenues for observing the sky.  The SKA has the potential to be a
highly multiplexed instrument, involving multi-beaming on a much
larger scale than can be envisioned with current telescopes, so that
multiple users can have access to the telescope simultaneously for
independent projects.  As computing power and memory become cheaper,
the SKA could be ``digitally upgraded,'' evolving into a more powerful
telescope with time.  Exploiting the possibility of rapid and large
disk and memory (RAM) storage of digital data, the SKA has the
potential to provide a ``look-back'' buffer, allowing the telescope to
be pointed in an \textit{ex post facto} manner based on a trigger,
say, from another telescope at another wavelength.  Finally, its
digital underpinnings naturally lead to rich contributions to Virtual
Observatories of the future.

\section{SKA Parameter Space}\label{sec:eou.parameter}

The SKA cannot explore much new
parameter space because angular resolution, spectral-temporal
resolution, and
polarization parameters have all been probed at most wavelengths with
the current generation of radio telescopes. However, the SKA will
greatly enlarge known parameter space by i)~a much greater
sensitivity, over a very wide range of angular resolutions; ii)~a
much larger instantaneous field-of-view at least at some wavelengths;
iii)~the potential for  multiple, independently steerable,
fields-of-view within which are independently steerable beams. The
sensitivity and sky coverage advances combine to provide two major
steps forward compared with current instruments i)~the volume of
space accessible to the SKA will be enormously increased, and hence
the chances of finding intrinsically rare objects in large scale
surveys will be much enhanced; ii)~the potential for ``all-sky''
coverage allows for statistical analysis of the surveys not biased by
the small area of sky being studied. A major advantage of multiple
independent beams is a multiplex one: several groups pursuing
different goals can operate on the SKA simultaneously and ``light'' will
thereby stream in through more windows on the Universe.

One intriguing expansion into new parameter space is negative time!  A
time buffer recording raw data (undetected voltages) would allow an
astronomer (or more likely a computer) to interrogate the raw SKA data
and form a full sensitivity beam anywhere in the field of view (FoV)
for periods of ten seconds to minutes earlier in time, hence the SKA
can have started monitoring \emph{before} receipt of a trigger.  This
trigger could be observations by the SKA itself or from some other
instrument, such as a satellite monitoring $\gamma$-ray bursts.

The second fundamental design driver of the SKA is the
instantaneous \hbox{FoV}.  Because radio wavelengths are 
long, a large FoV is natural but changing technology may enable even
greater increases. The SKA design goal has specified an FoV of at least
1~deg${}^2$ at a wavelength of~21~cm. In some concepts being
pursued, this may rise to as much as 100~deg${}^2$ or more.

While the instantaneous FoV that might be achieved for the SKA is
impressive ($> 10$~deg${}^2$) when compared to modern radio
telescopes, it is smaller (in some cases much smaller) than some
historical radio telescopes.  The SKA's potential to make a
qualitative improvement over previous and existing telescopes is its
simultaneous combination of large instantaneous FoV with high
sensitivity and angular resolution.  Historically, radio telescopes
have been capable of obtaining either large solid angle coverage
(e.g., the STARE survey at~610~MHz by Katz et al.~\cite{khcm03} with a
FWHM beam of~4000\arcmin) or high sensitivity (e.g., the Arecibo
telescope with a gain of~11~K~Jy${}^{-1}$ at~430~MHz) but not the two
simultaneously.  If the SKA satisfies only its design requirements, it
will produce at least an order of magnitude sensitivity improvement,
over a large frequency range, with a solid angle coverage that is
comparable to or exceeds that of all but a small number of
low-sensitivity telescopes.

The SKA's potential is even more striking given that all of the Key
Science Projects identified for the SKA project involve surveys of one
sort or another.  The surveying speed of a telescope system to reach a
given flux density limit is proportional to the product of its
instantaneous FoV and the square of its sensitivity. In
Table~\ref{tab:eou.speed} we compare the relative continuum surveying
speeds of current instruments with that of the \hbox{SKA}. It is clear
that the SKA will have a surveying speed at least 10,000 times greater
than is possible with the current instruments. If the advantages of
the wide FoV concepts being investigated can be realised, then a
factor larger than $10^5$ can be achieved. New discoveries often arise
from large-scale surveys during which the rare discrete objects of a
new type or new large-scale emergent properties of the ensemble of
objects stand out. The advance offered by the SKA promises a complete
revolution in our ability to survey the radio sky---so far in advance
of anything which has been done before that it truly can claim to take
the SKA into an arena of ``discovery science.''

\begin{table}
\caption{Relative Surveying Speed for Continuum
	Sources\label{tab:eou.speed}}
{\small
\begin{tabular}{lccc}
\noalign{\hrule}
          & Relative    & Field         & Survey \\
Telescope & Sensitivity & of View       & Speed \\
          & (per beam)  & (deg${}^2$)   & \\
\noalign{\hrule\hrule}

SKA specification & 1      & 1     & 1 \\
$\;$($A_{\mathrm{eff}}/T_{\mathrm{sys}}=20,000$) & & & \\

SKA potential     & 1      & 100   & 100 \\

Arecibo/ALFA      & 0.038  & 0.03 & $10^{-4.4}$ \\
%   (200m, 25K)
$\;$($A_{\mathrm{eff}}/T_{\mathrm{sys}}=760$) & & & \\

VLA               & 0.011  & 0.26  & $10^{-4.5}$ \\
%  (130m, 35K)
$\;$($A_{\mathrm{eff}}/T_{\mathrm{sys}}=220$) & & & \\

Parkes            & 0.004 & 0.52  & $10^{-5.1}$ \\
% (64m, 25K, 13 beam)
$\;$($A_{\mathrm{eff}}/T_{\mathrm{sys}}=80$) & & & \\

GBT/Effelsberg    & 0.012 & 0.02 & $10^{-5.7}$ \\
%  (100m, 25K)
$\;$($A_{\mathrm{eff}}/T_{\mathrm{sys}}=240$) & & & \\
\noalign{\hrule}
\end{tabular}
}
The tabulated surveying speeds are applicable to observations at a
wavelength of~21~cm.
The surveying speed is calculated as the square of the sensitivity
multiplied by the instantaneous \hbox{FoV}. In calculating the
relative sensitivity per beam, all continuum bandwidths have been
assumed to be equal.
\end{table}

\section{New Opportunities for the SKA}\label{sec:eou.science}

Although the SKA will not investigate new dimensions in parameter
space per se, its capabilities will allow new opportunities for
investigating the radio sky.  Here we summarize two of the avenues
that seem most promising today, recognizing that other possibilities
may arise in the next 20--50~years.

\subsection{The Dynamic Radio Sky}\label{sec:eou.dynamic}

A combination of high time resolution, high sensitivity, and large
instantaneous FoV will enable the SKA to search for and study
transient sources. Although various classes of radio transients are
known today---such as giant pulses from radio pulsars, radio flares
from micro-quasars and brown dwarfs, maser flares, supernovae events,
and radio afterglows from $\gamma$-ray bursts---except for pulsars,
there have been few surveys of a large area of sky at radio
wavelengths for transient sources.

\begin{figure}[tbh]
\includegraphics[width=\columnwidth]{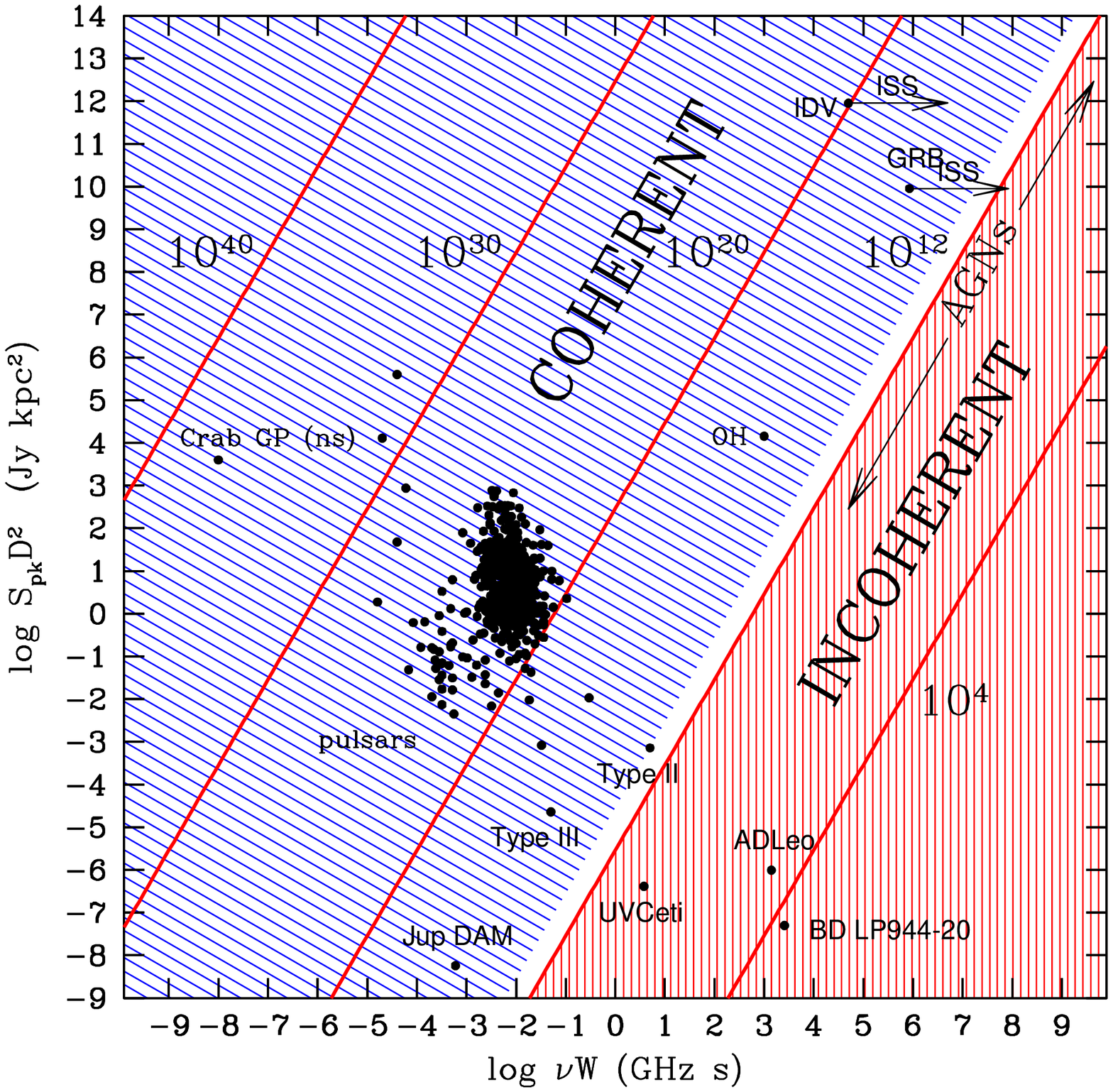}
\vspace{-1.5cm}
\caption[]{Phase space for known and anticipated transient signals.
The horizontal axis is the product of transient duration~$W$ and radio
frequency~$\nu$ while the vertical axis is the product of the peak
flux density~$S_{\mathrm{pk}}$ and the square of the distance~$D$ and
is proportional to luminosity.  Lines of constant brightness
temperature are shown, and the uncertainty principle requires that
signals be to the right of $\nu W = 10^{-9}$.  Also shown are  representative examples of active
galactic nuclei (AGNs); $\gamma$-ray burst afterglows (GRBs); pulsars,
including the 2~ns giant pulses from the Crab pulsar (Crab~GP~[ns], \cite{hkwe03});
flare stars and brown dwarfs; masers; planets; and the Sun.  In
addition, propagation effects produce intraday variability (IDV) and
interstellar scintillation (ISS), which modulate compact radio sources
on a variety of time scales.  These effects make sources appear to be
more variable than they are intrinsically but are rich in information
about the intervening media.  The known objects span nanosecond to
year durations.  The dynamic radio sky represents an opportunity for
the SKA to expand our knowledge of the radio sky in the same way that
high-energy transient studies, conducted with wide-field X- and
$\gamma$-ray instruments, have discovered $\gamma$-ray bursts,
accreting black holes and other compact objects, and flares from
planets, stars and AGNs.}
\label{fig:eou.transient}
\end{figure}

The success of X- and $\gamma$-ray telescopes in carrying out such
wide-field surveys suggests that wide-field radio surveys would be
equally fruitful.  Figure~\ref{fig:eou.transient} illustrates the
potential range of radio transient phenomena is large.  In the
Rayleigh-Jeans approximation, a source with brightness temperature~$T$
varies (intrinsically) on a time scale or pulse width~$W$,
\begin{equation}
W^2 = \frac{1}{2\pi k}\frac{SD^2}{T}\frac{1}{\nu^2},
\label{eqn:eou.phasespace}
\end{equation}
where the observed flux density is $S$, the source's distance is $D$,
the emission frequency is $\nu$, and~$k$ is Boltzmann's constant.  As
Figure~\ref{fig:eou.transient} shows, the range of $\nu W$ covers at
least 13 orders of magnitude while the range of $SD^2$ covers at least
20 orders of magnitude.  Moreover, in addition to the known classes of
radio transients, a variety of other classes involving either coherent
or incoherent processes have been suggested such as
\begin{itemize}
\item Radio flares from extra-solar giant planets, akin to Jovian
decametric radiation;
\item Prompt emission from supernovae, as in SN~1987A; 
\item Radio-loud, $\gamma$-ray quiet ``$\gamma$-ray bursts'';
\item Prompt emission from $\gamma$-ray bursts;
\item Radio hypernovae associated with Population~III stars;
\item Galactic flares, outbursts due to the tidal disruption of
single stars falling into the supermassive black holes at the centre
of apparently normal galaxies;
\item Evaporating black holes; and
\item Extraterrestrial transmitters.
\end{itemize}

\subsection{Synergy with Other Telescopes}\label{sec:eou.synergy}

The sensitivity of the SKA means that for the first time many objects
which dominate the science done at other wavebands (stars, galaxies,
star forming regions, proto-galaxies, galaxy clusters) become readily
observed at radio wavelengths. The SKA will be a leading member of a
complementary group of next-generation photon collectors including:
ground-based optical telescopes in the 30--50~m class; the JWST (near-
and mid-IR); ALMA (mm- and sub-mm wave); next-generation X-ray and
$\gamma$-ray observatories (XEUS and Constellation-X). All of these
will provide imaging information on the scale of roughly 0.1~arcsecond
or better and all will provide unique views of the
Universe. Completely new information will then flow from a comparison
of the enormous data sets which will be produced from synoptic surveys
in many wavebands. From a combination of these surveys, statistically
significant patterns and subtle correlations between parameters will
become apparent, pointing to new phenomena and rare or previously
unknown objects will stand out. Meanwhile the first generations of
neutrino detectors, the ground-based gravitational wave observatories
(LIGO; GEO-600; VIRGO, and others) and the space-based follow-ons
(LISA) are expected to lead to the detection of gravitational waves
from a variety of sources. All of these ventures and their anticipated
outcomes in terms of discovery and understanding will benefit from
observations in the radio band. The SKA will provide the sensitivity
along with a rich set of observational modes to enable exploration of
the radio sky in all of its anticipated complexity.

\section{Design Philosophies for the SKA}\label{sec:eou.design}

In designing the \hbox{SKA}, we need to look beyond an instrument
which will merely satisfy the requirements of the current Key Science
Projects---challenging though these are---and envisage an instrument
which also allows new observing paradigms. Some guiding principles are
\begin{itemize}
\item Seek the maximum flexibility in the design, to enable
the combination of high sensitivity simultaneously with high
resolution in time, frequency and/or angle.  Where multiple
requirements cannot be met simultaneously, allow the ability to make
trade off between different aspects, e.g., collecting area against
solid-angle coverage, or signal processing bandwidth at high
frequencies against FoV at low frequencies.

\item Seek ways for the
multiple reuse of expensive components, for example via simultaneous
multiple beams or spectral bands.

\item Exploit the areas of fastest
changes in technology by designing the signal processing hardware in
such a way that the major purchases can be made late in the project.

\item Plan the architecture for upgrades right from the start by adopting
an open modular approach to the various systems and sub-systems. In
particular digital signal processing and computing power will increase
by many orders of magnitude while decreasing in cost over the course
of the SKA's lifetime.  This will allow us to generate more beams and
hence more channels through which to examine the sky. The SKA's
architecture should recognize this and provide clear interfaces,
down-stream of which it is relatively easy to upgrade the electronics
in a step-wise fashion.

\item Begin the software development early; the
SKA will inevitably be the largest software project in astronomy, and
there is no Moore's Law escalator to help.

\item Facilitate time-buffering
and archiving as much of the raw data as possible to enable the
pre-trigger beam-forming mode discussed in~\S\ref{sec:eou.technology}
\end{itemize}

Ideally the
SKA will be designed so as to provide independently-steerable
fields-of-view and independently-steerable beams within each
\hbox{FoV}, over at least the lower portion of its operational
frequency range. Such architecture would provide users with a highly
flexible and responsive instrument and would inevitably lead to
changes in observing style. It remains to be seen how feasible it is
to achieve all this flexibility from the start. However, we note that such
an approach generically incorporates
\begin{itemize}
\item A science survey advantage, which is required for a range of science
programmes requiring huge amounts of telescope time and which would be impossible with conventional
systems;
\item Long term monitoring opportunities with dedicated beams;
\item A ``community'' advantage because many groups could access the whole aperture
simultaneously, allowing the operation of the SKA to resemble that of
particle accelerators or synchrotron light sources;
\item A multiplex advantage, simply by increasing the range of
different data sets which can be collected; and
\item An adaptive beam-forming advantage, as ``reception nulls'' could
be steered to cancel out sources of \hbox{RFI}.
\end{itemize}

In addition to the Key Science Projects, the different FoVs could, for
example, be used for
\begin{itemize}
\item Monitoring variable sources such as looking for pulsar glitches
and timing changes due to gravitational radiation or observing effects
of interstellar scintillation;

\item Integrating for long periods of time to obtain  the ultimate in
sensitivity;

\item Studies of time variable phenomena on time scales from
nanoseconds and longer and seeking transient radio sources and
responding instantly to transients discovered in other wavebands; or

\item Experimentation or ``high risk'' observations which would
not be scheduled by the standard peer-review process.
\end{itemize}

\section{Exploration-Driven Observing}\label{sec:eou.modes}

The SKA will be sufficiently different from current instruments that
the ``standard model,'' discussed in \S\ref{sec:eou.intro}, via which
the user interacts with the instrument and its data, can be
re-evaluated. It is vital that the flexible modes of use of the SKA
not become too formalised. Many astronomers are conscious that the
standard model is not the way to make breakthroughs and other
responses are already being tried. The NVSS and FIRST surveys and the
Hubble Deep Field are examples of one different approach where some of
the time on a telescope is devoted to large programs whose data
products are then made available publicly. The success of these
programmes encourages us to look for further ways to break the
constraints of the traditional common-user paradigm and an excess of
egalitarianism.

Even the
existing science goals for the SKA demand a great deal of flexibility
in the way the total collecting area,  frequency coverage, and angular
resolution can be exploited.  For example, some observations require
high resolution in time, frequency, and/or angle along with the ability
to trade off collecting area against solid-angle coverage. For the SKA
we wish to use the sensitivity for both imaging and non-imaging
science in ways that adopt the best features of existing arrays and
filled aperture antennas.

Modes of operation carry with them implications for data processing
and analysis. Data rates are unprecedented and will require large
increases in computing capacity and new algorithms over the next
10--15~years and beyond.  Brute force solutions (all of everything)
will not be practical so trade-offs will have to be made.  Flexible
implementations of the various modes will allow these trade-offs
between resources while maximizing the scientific value of the
observations, but when searching for the unknown we must avoid making
the observational filter so narrow that we see only what we already
know to exist.

Particular modes of observation that need to be enabled in order to
address the Key Science Projects and especially for the exploration of
as-yet unknown phenomena include

\begin{description}
\item[Full Field-of-view Imaging]
This mode will be the one used by most of the surveys in the Key
Science Projects.  Enabling of this mode requires channel widths and
correlator dump times that scale as the reciprocal of the maximum
baseline.  In order to avoid excessive signal transport and processing
loads, this mode would normally be limited to the central part of the array.

\item[High-resolution Imaging]
With the \hbox{SKA}, real
time correlation for high-resolution imaging will be possible with
data transport by fibre.  However, data rates will not allow full FoV
imaging using the longest baselines in the envisioned array.  Smaller
regions can be processed at the full resolution, and it should be also
possible to image multiple smaller patches, e.g., around each radio
source in the \hbox{FoV}.

\item[Planetary Radar/Time-Gated Imaging]
High resolution imaging of radar return signals can remove the
ambiguities usually associated with delay-Doppler radio imaging can be
removed by forming delay-Doppler images within each resolution element
of a synthesized image.  Similar gating can be used to analyse other
periodic signals, e.g., to ``gate out'' pulsars in a search for
un-pulsed emission.

\item[Slow Transient Searches]
Slow transients are defined as those with durations long enough that
they can be found or studied by moving the FoV (e.g., a raster scan)
over the region being studied.  For GRB afterglows, for example, the
entire available sky could be sampled in~1~day using a mosaic imaging
mode that provides roughly 2~s of integration per direction even for a
1~deg${}^2$ \hbox{FoV}.

\item[Fast Transient Searches]
Fast transients require a long-dwell staring mode.  For wide-field
blind surveys, the instantaneous solid angle needs to be maximized.
This places a limit on maximum baseline that can be used in order to
keep data processing rates manageable.  The fastest transients, e.g.,
$\Delta t \le 1$~s, are subject to distortion by dispersive
propagation like that seen from pulsars and therefore require adequate
frequency resolution for ``dedispersion'' methods to be applied.

\item[Blind Pulsar Surveys]
For pulsars, which are essentially a special case of fast transients,
it will be possible to analyse only the inner core array of size
$b_{\mathrm{max}} \sim 1$ to~5~km.  Even so, this already implies
about~$10^4$--$10^5$ pixels and the analysis for each pixel includes
summing over frequency using roughly $10^3$ trial dispersion measure
values, followed by Fourier analysis of each resulting time series and
statistical threshold tests.  In order to find pulsars in compact
binaries with orbital periods less than a few hours, a search over an
acceleration parameter must also be done for a few hundred values.

\item[Blind SETI Surveys]
SETI usually
involves searches for possibly modulated, very narrow-band carrier
signals.  Interstellar propagation limits the minimum signal bandwidth
to about~0.05~Hz for distances larger than about~300~pc at~1~GHz.
Spectra must therefore be computed with billions of channels and
concordant time resolution for each pixel.  As with pulsars, blind
surveys will use the inner core array.

\item[Targeted Observations]
When a
source's location is known or constrained from other astronomical
observations, processing requirements are lessened significantly and
can exploit real-time beamforming capabilities.  With multiple beams
within the primary \hbox{FoV}, searches can be made on multiple
targets. Analyses of signals from each beam may require
special-purpose hardware (SETI spectrometers, pulsar-timing machines,
etc.) to handle the data rates.

\item[Incoherent Summing Modes]
When the full gain of the SKA is not needed, but access to the entire
primary FoV is desired, signals from antennas can be summed
incoherently.  For $N$ antennas divided into $N_{\mathrm{sa}}$
subarrays, the sensitivity per analyzed subarray signal is
$(G/T)_{\mathrm{sa}} = (G/T)_{\mathrm{SKA}}/(N \times
N_{\mathrm{sa}})^{1/2}$.  With this mode, for example, 100 subarrays
could yield instantaneous coverage of~100~deg${}^2$ and the
sensitivity equivalent of a 30-m antenna.  This mode can provide very
high time resolution without requiring a very high data processing
rate and may find use with the SKA early on in its deployment.

\item[RFI Mitigation Modes]
Data rates are also entwined with issues pertaining to shared use of
the radio spectrum, which is likely to grow worse with time.
Specialized data acquisition modes for pulsars and transients, with
high resolution in time and frequency, may also be necessary to ensure
that RFI mitigation algorithms, which might otherwise assume that
interfering signals fill only fractions of the overall time-frequency
plane, do not excise the very signal being sought.  Clearly sites with
low levels of terrestrial interference are preferred in order to
maximise the probabilities of recognising unexpected types of signals.
\end{description}

\section{The Human Factor}\label{sec:eou.human}

\begin{quote}
\emph{``In the field of observation, chance favours the prepared
mind'': L.~Pasteur}
\end{quote}

\begin{quote}
\emph{``The harder I practice, the luckier I get'':  Sports adage}
\end{quote}

So far we have addressed
essentially technological solutions to the question, How do we make
the SKA perform as a discovery instrument?  These can all be gathered
together within the generic answer, Provide access to new regions of
a multi-parameter space via the application of innovative radio
technology.  There is more to discovery than can be plotted as
parameter space on a graph, though---there is an abstract or innovation aspect
to discovery as well, which is just as important but is often
overlooked. We should be inspired by another lesson of history.

\begin{quote}
\emph{Astronomical discoveries are usually made by people who are
``curious'' and take the time to understand their instrument, with
less emphasis on a quick publication.}
\end{quote}

The lesson is to allow people enough ``room''
to make discoveries. Discoveries invariably result from an individual
becoming so familiar with the data, \emph{and hence the possible sources of
error in them}, that he/she can recognize and unexpected clue for what
it is worth. The discovery of pulsars is a perfect example. It was
only the apparently tedious day-to-day routine of checking miles of
pen-recorder charts that enabled Jocelyn Bell to spot her famous ``bits
of scruff'' and to distinguish them from the usual types of
interference. She paid her dues and for the right mind familiarity
does not breed contempt for the data. The discovery of Jupiter bursts,
interplanetary scintillations, the CMB, and even the original
discovery of Jansky himself, are other examples.  While we cannot
control the subtleties of individual human curiosity, we can provide
the circumstances in which it can flourish.

The continued success
stories of pulsar groups around the world, stretching over nearly four
decades in time, provides further lessons for SKA planners. In
addition to choosing a field which is phenomenologically rich and for
which the inverse problem, connecting data to fundamental physics, is
tractable, what else have they done right?
\begin{itemize}
\item Designed their own ``back-end'' equipment and software and
constantly updated it.
\item Archived the raw data and re-analysed them as computing power
increased.
\end{itemize}
Some possible responses to these lessons for SKA planners
are
\begin{itemize}
\item Award some time to successful groups or collaborations on the
basis of their past record---a ``rolling time allocation grant'' which
can be sustained or closed down on the basis of performance integrated
over several years.
\item Allow high-risk or unproven new-style
observations. The availability of independent beams will help to make
this feasible without compromising conventional observing programs.

\item Build an open system that allows user-produced hardware and software
to be employed on the telescope. Clean interfaces will need to be
defined and maintained.

\item Maintain the technical expertise in community. The experience in
the USA over the past two decades during which technology development
in many of the traditional university departments systematically fell
away as resources were directed to the national facilities, should act
as a warning.  In the lead up to the SKA innovative R\&D is
re-invigorating the world-wide community and involving a new
generation of engineers and students.  When the SKA is completed, it
is vital to allow a cadre of technical people world-wide to gain
continuous access to parts of the system for continuous
experimentation.
\end{itemize}

\section{Summary}\label{sec:eou.conclude}

The impact of the SKA will depend not so much on the cleverness of the
astronomers in defining the important astrophysical problems of today,
but in the cleverness and ambition of its designers to obtain better
sensitivity, higher dynamic range, larger field-of-view, and perhaps
some other parameters not yet contemplated.

The SKA planners and designers must therefore
have a vision for a system that is not only much more powerful than
previous radio telescopes but at the same time is highly flexible,
easy-to-use, and has an operating philosophy which positively
encourages and allows the astronomers of tomorrow to look at the sky
and to examine their  data in new and creative ways. It must also
allow the radio engineers of tomorrow the space to conceive and design
the innovative upgrades which increased signal processing and
computing power will allow.

Just as important will be to attract and nourish talented and careful
observers who can become so familiar with particular types of data as
to be able to spot the important new clues which will lead to fresh
discoveries. The SKA will need to solve their problems, not our
problems, which they will have already solved or shown to be naive or
irrelevant. Including ``Exploration of the Unknown'' as one of the
prime goals of a highly-flexible SKA is founded firmly on the profound
contributions which radio astronomy has already made to our knowledge
of the Universe.

\end{document}